\documentstyle[epsf,12pt]{article}
\def\mevc {\ifmmode {\rm MeV}/c \else MeV$/c$\fi}
\def\mevcc {\ifmmode {\rm MeV}/c^2 \else MeV$/c^2$\fi}
\def\gevc {\ifmmode {\rm GeV}/c \else GeV$/c$\fi}
\def\gevcc {\ifmmode {\rm GeV}/c^2 \else GeV$/c^2$\fi}
\def\ra   {\rightarrow}
\newcommand{\Pt} {\ifmmode p_{\rm t} \else $p_{\rm t}$\fi}
\newcommand{\Et} {\ifmmode E_{\rm t} \else $E_{\rm t}$\fi}
\newcommand{\met} {\ifmmode \not\!\!E_{\rm t} \else 
                \mbox{$\not\!\!E_{\rm t}$}\fi}
\newcommand{\sigtt} {\ifmmode \sigma_{t\bar t} \else $\sigma_{t\bar t}$\fi}
\newcommand{\mtop} {\ifmmode m_{\rm top} \else $m_{\rm top}$\fi}
\newcommand{\lxy} {\ifmmode L_{\rm xy} \else $L_{\rm xy}$\fi}
\newcommand{\Bs} {\ifmmode B_{\mbox{\sl s}}^{0}
                       \else $B_{\mbox{\sl s}}^{0}$\fi}
\newcommand{\Ds} {\ifmmode D_{\mbox{\sl s}}^{-}
                       \else $D_{\mbox{\sl s}}^{-}$\fi}
\newcommand{\Bsh} {\ifmmode B_{\mbox{\sl s}}^H
                       \else $B_{\mbox{\sl s}}^H$\fi}
\newcommand{\Bsl} {\ifmmode B_{\mbox{\sl s}}^L
                       \else $B_{\mbox{\sl s}}^L$\fi}
\newcommand{\Dsl} {\ifmmode D_{\mbox{\sl s}}^{-} \ell^+
                       \else $D_{\mbox{\sl s}}^{-} \ell^+$\fi}
\newcommand{\ctau} {\ifmmode c\tau \else $c\tau$\fi}
\newcommand{\phipi} {\ifmmode \phi \pi^-
                       \else $\phi \pi^-$\fi}
\newcommand{\kstark} {\ifmmode K^{*0} K^-
                       \else  $K^{*0} K^-$\fi}
\newcommand{\ksk} {\ifmmode K^0_S K^-
                      \else $K^0_S K^-$\fi}
\newcommand{\phil} {\ifmmode \phi \mu^- \nu
                      \else $\phi \mu^- \nu$\fi}
\newcommand{\Dsmu} {\ifmmode D_{\mbox{\sl s}}^{-} \mu^+ 
                       \else$D_{\mbox{\sl s}}^{-} \mu^+$\fi}
\newcommand{\dgam} {\ifmmode \Delta\Gamma \else $\Delta\Gamma$\fi}
\newcommand{\dgog} {\ifmmode \Delta\Gamma/\Gamma \else 
                            $\Delta\Gamma/\Gamma$\fi}
\newcommand{\dm} {\ifmmode \Delta m \else $\Delta m$\fi}
\newcommand{\dms} {\ifmmode \Delta m_{\mbox{\sl s}} \else 
                           $\Delta m_{\mbox{\sl s}}$\fi}
\newcommand{\ptrel} {\ifmmode p_{\rm t}^{\rm rel} 
                       \else $p_{\rm t}^{\rm rel}$\fi}
\newcommand{\ed} {\ifmmode \varepsilon D^2 \else $\varepsilon D^2$\fi}
\begin{document}
{\Large\bf Heavy Quarks at the Tevatron: Top \& Bottom}
\\[2mm]
{\it Manfred Paulini 
(Representing the CDF and D\O\ Collaboration)} \footnote[4]{To 
appear in the
Proceedings of IVth International Workshop on Progress in Heavy Quark
Physics, Rostock, Germany, 20-22 September 1997.}
\\[2mm]
{\small Lawrence Berkeley National Laboratory, Berkeley, CA 94720, USA}

\begin{abstract}
We review the status of heavy quark physics at the Fermilab Tevatron
collider by summarizing recent top quark and $B$~physics
results from CDF and D\O. In particular, we discuss the measurement of
the top quark mass and top production cross section as well as 
$B$ meson lifetimes 
and time dependent $B^0\bar B^0$ mixing results.
\end{abstract}

\section{Introduction}

In this report, we review recent results on heavy quark physics (top
\& bottom) from
the Tevatron $p\bar p$~collider at Fermilab
operating at a centre-of-mass energy of $\sqrt{s} = 1.8$~TeV. 
During the Tevatron Run~I, which ended in 1996, the CDF
experiment~\cite{cdf_det} and the 
D\O\ detector~\cite{d0_det} recorded each 
an integrated luminosity of about 100~pb$^{-1}$.

In Section~\ref{top_sec},
we summarize the status of top quark physics at
CDF and D\O\ discussing 
measurements of the top production cross section and
the top quark mass.
Section~\ref{b_sec} is devoted to
recent $B$ physics results from the Tevatron collider, where we concentrate
on $B$ hadron lifetimes and time dependent $B^0\bar B^0$ mixing
results. 
We conclude with Section~\ref{conclude_sec}.       

\section{Top Quark Physics at the Tevatron}
\label{top_sec}

At the Tevatron the dominant top
quark production mechanism is $t\bar t$ pair production through $q\bar
q$ annihilation. Gluon-gluon fusion contributes 
to the $t\bar t$ cross section 
to about 10\% at $\sqrt{s} = 1.8$~TeV. 
During the Tevatron Run~I over $5\cdot10^{12}$ $p\bar p$ collisions
took place within the CDF and D\O\ detectors but only about 500 $t\bar t$
pairs have been produced per experiment implying
\sigtt\ to be 
about ten orders of magnitudes lower than the total
inelastic cross section. 
This means
the challenge in studying top quarks is to
separate them from backgrounds in hadronic collisions.  

Within the Standard Model, each of the pair produced top quarks decays
almost exclusively 
into a $W$ boson and a $b$ quark. The
$W$ boson decays into either a lepton-neutrino or 
quark-antiquark pair. The top decay signature depends primarily on the
decay of the $W$ boson. Events are classified by the number of
$W$'s that decay leptonically.

\begin{figure}[tb]
\begin{picture}(160,45)(0,0)
\put(-3,36){\large\bf (a)}
\put(53,36){\large\bf (b)}
\put(109,39){\large\bf (c)}
\centerline{
\epsfysize=5.0cm
\epsffile{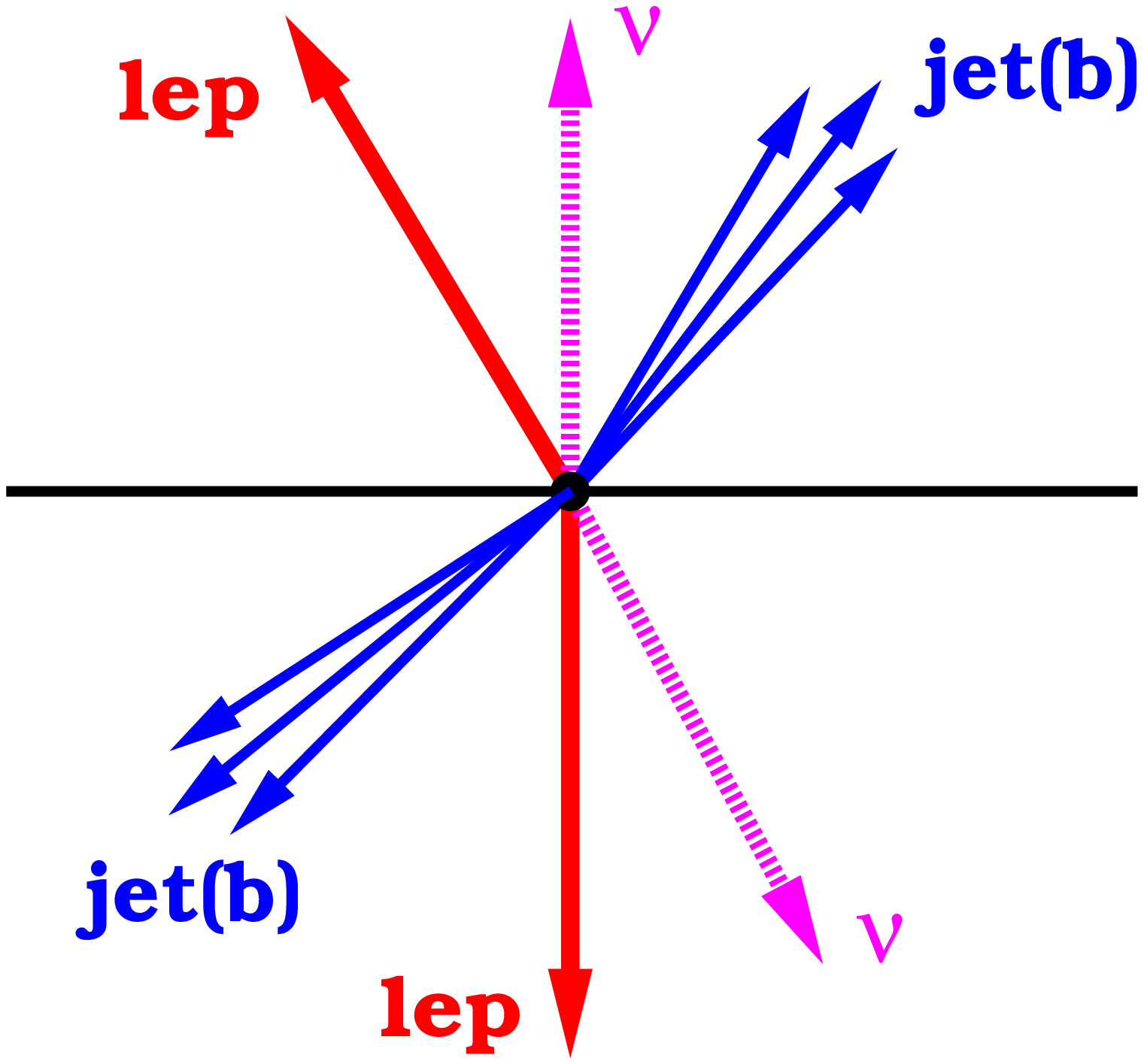}
\hspace*{0.2cm} 
\epsfysize=5.0cm
\epsffile{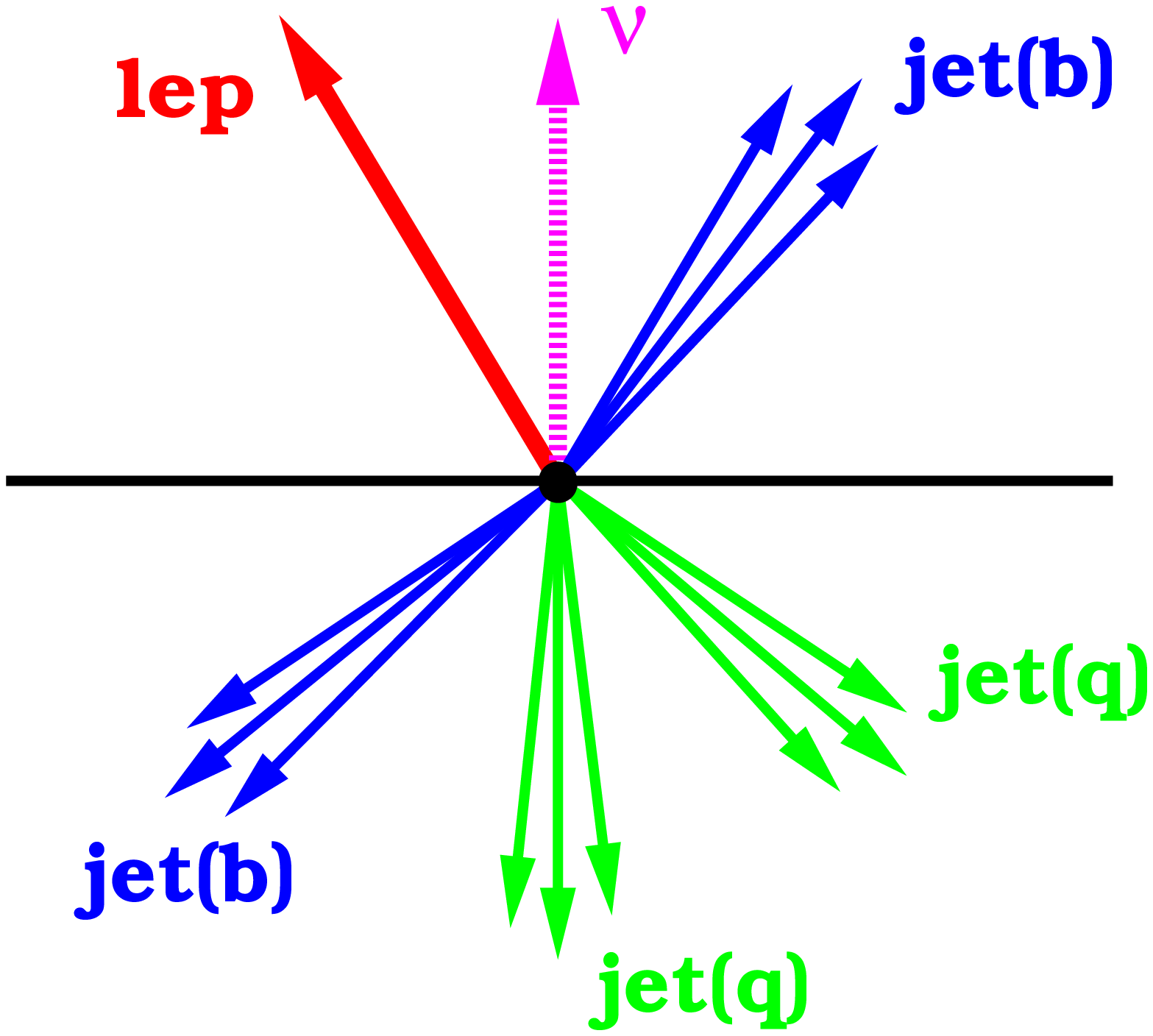}
\epsfysize=5.0cm
\epsffile{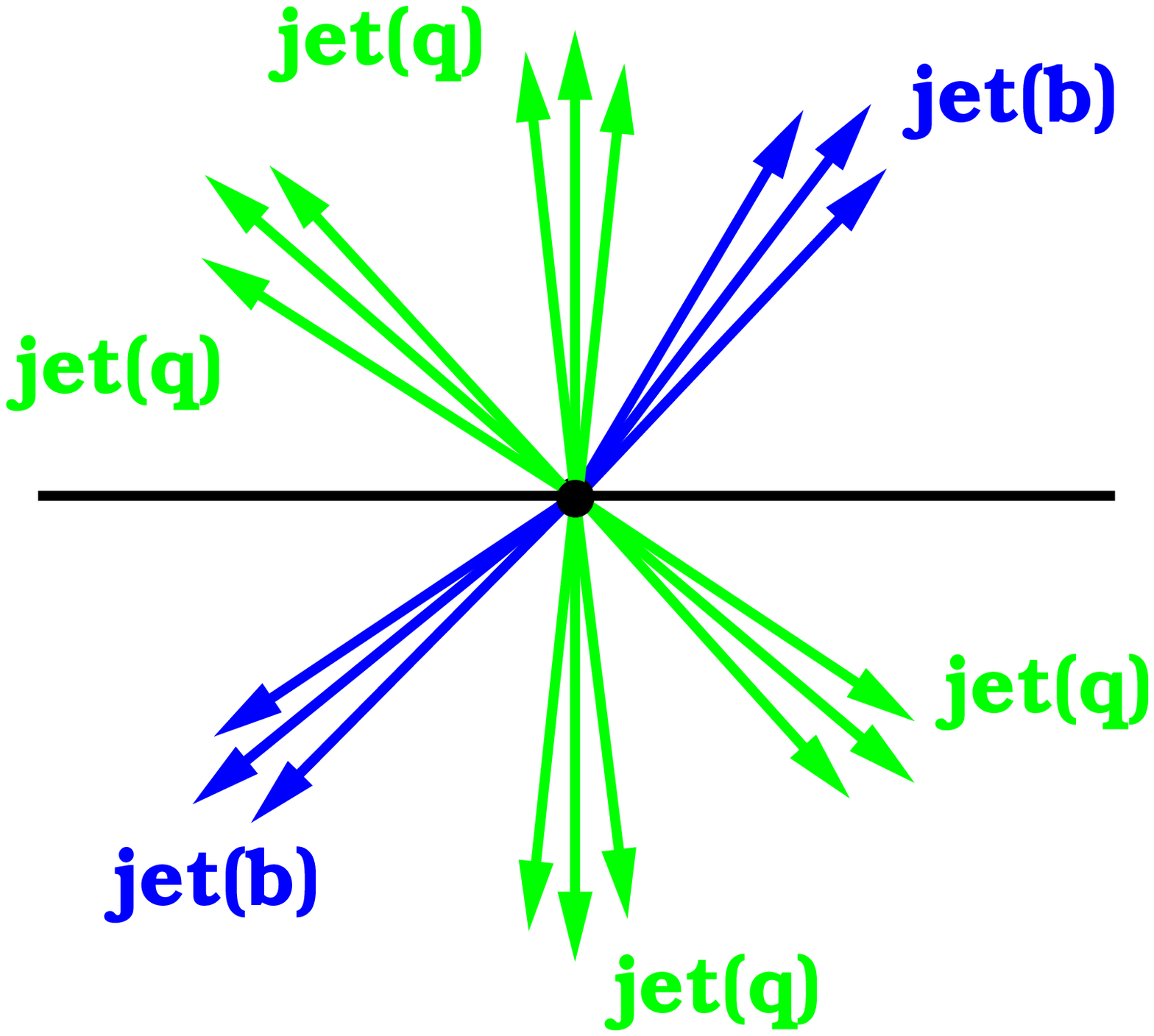}
}
\end{picture}
\caption{Top decay signature of (a) the dilepton channel, (b) the
lepton plus jets channel, and (c) the all hadronic channel.}
\label{top_dec_sig1}
\end{figure}

If both $W$ bosons decay leptonically into $W \ra \ell\nu$, where
lepton $\ell$ refers to $e$ or $\mu$, we call it the
`dilepton channel'. The final state consists of
$\ell^- \bar\nu \ell^+ \nu b\bar b$ as can be seen in
Figure~\ref{top_dec_sig1}a). Due to both $W$'s
decaying semileptonically, this top decay mode has a small branching
fraction of about 5\%. 
If one of the $W$ bosons decays leptonically into $W \ra \ell\nu$ and
the other into $W \ra q\bar q^{\prime}$, we call it the
`lepton plus jets channel', where the final state consists of
$\ell \nu q\bar q^{\prime} b\bar b$ as shown in
Figure~\ref{top_dec_sig1}b). 
This decay mode happens with a branching ratio of about 30\%. 
If both $W$ bosons decay into quark pairs as $W \ra q\bar q^{\prime}$,
we call it the `all hadronic channel'. The final state consists of
$q\bar q^{\prime} q\bar q^{\prime} b\bar b$ as can be seen in
Fig.~\ref{top_dec_sig1}c).  
This top decay mode occurs at a large rate of about 44\%. The
remaining top decays involve a $\tau$~lepton plus another
lepton (6\%) or a $\tau$~lepton plus jets (15\%).

\subsection{The Counting Experiments}

\subsubsection{The Top Dilepton Channel}
 
The signature of the dilepton channel (see Fig.~\ref{top_dec_sig1}a)
is two isolated high \Pt\ leptons and missing energy 
(\met) from the two neutrinos that escape the detector
unobserved. In addition, there are two jets 
originating from $b$ quarks in the
event. After demanding two leptons and \met, the event selection also
relies on kinematic requirements. 
The dominant backgrounds are from 
Drell-Yan production, $Z \ra \tau\tau$, fake leptons, and
$WW$ diboson
production.
The dilepton channel has a good signal to background ratio, but low
statistics. The dilepton event summary for
CDF and D\O\ is shown in Table~\ref{dilep_sum} (left) and
further described in Ref.~\cite{D0_dilep,CDF_dilep}.
Both experiments find a few events on small backgrounds.
Table~\ref{dilep_sum} also contains the D\O\
$e\nu$ channel~\cite{D0_top_prod} and the CDF $e$ or $\mu$ plus $\tau$ mode~\cite{CDF_top_tau}. 

\begin{table}[bt]
\begin{center}
\vspace*{-0.6cm}
  \begin{tabular}{|l||c|c||l||c|c|}
   \hline 
   Dilepton    &    D\O      &   CDF  &
   Lepton plus jets    &    D\O     &   CDF   \\
   \hline 
   \hline 
  \underline{$e \mu$:}   &     3         &   7      &
  \underline{Event shape:}   &   19       &   22      \\
   Background  &   $0.2\pm0.2$        &  $0.8\pm0.2$       &
   Background   &    $8.7\pm1.7$       &  $7.2\pm2.1$        \\
   Expected yield  &  $2.2\pm0.5$ & $2.6\pm0.2$ &
   Expected yield  &   $14.1\pm3.1$  &   (67 pb$^{-1}$) \\
    \hline
  \underline{$e e$ or $\mu \mu$:}   &    2        &   2      &
  \underline{Lepton tag:}    &    11      &  40       \\
   Background  &   $1.2\pm0.3$        &   $1.6\pm0.4$      &
   Background  &    $2.4\pm0.5$    &  $24.3\pm3.5$      \\
   Expected yield  &  $1.9\pm0.3$  & $1.8\pm0.2$ &
   Expected yield &   $5.8\pm1.0$  &  $9.6\pm1.7$  \\
    \hline
  \underline{$e\nu$ (D\O), $\ell\tau$ (CDF):}   &    4       &    4     &
  \underline{SVX tag:}   &    -      &    34     \\
   Background  &     $1.2\pm0.4$       &  $2.0\pm0.4$       &
   Background  &    -      &  8.0 $\pm$ 1.4       \\
   Expected yield  & $1.7\pm0.5$     &   $0.7\pm0.1$ &
   Expected yield & - &   $19.8\pm4.0$ \\
   \hline
\end{tabular}
\caption{Event summary for the dilepton channel (left) and the lepton
	plus jets channel (right).   
	The expected yield from $t\bar t$ production is based on
	determinations of the  
	top cross section \cite{laenen} for a top quark mass of
	170~\gevcc\ (D\O) and 175~\gevcc\ (CDF), respectively.}
\label{dilep_sum} 
\end{center}
\vspace*{-0.5cm}
\end{table}

\subsubsection{The Top Lepton plus Jets Channel}
\label{toplepjet}
 
The signature of the lepton plus jets channel (see Fig.~\ref{top_dec_sig1}b)
is one isolated high \Pt\ lepton, missing energy 
(\met) from the neutrino and four jets where two of them are from $b$ quarks.
The dominant background is from $W$ plus jet production, where the
jets tend to be softer than jets 
in $t\bar t$ events. In addition, $t\bar t$ events always contain $b$ quarks. 
Both experiments follow different strategies to reduce
the background in this channel.

CDF tags the $b$ jets in the event through a
`soft lepton tag' (SLT) and a `SVX tag'. The first technique
identifies $b$ jets by searching for typically low momentum leptons
from $b \ra \ell X$ or  
$b \ra c \ra \ell X$ decays.
Electrons and muons are found 
by matching tracks from the central
drift chamber with
electromagnetic energy clusters in the calorimeter or track segments in the   
muon chambers. 
The efficiency for SLT tagging a $t\bar t$ event is $(18\pm2)\%$ with
a typical mistag rate per jet of about 2\%.
Details of the SLT algorithm
can be found in Ref.~\cite{CDF_top_prd}.
The second, more powerful $b$ tagging technique (SVX tag) exploits the finite
lifetime of $b$ hadrons by searching for a secondary decay vertex
displaced from the primary event vertex with
CDF's silicon vertex detector. The efficiency for SVX tagging a $t\bar t$
event is $(39\pm3)\%$, while the mistag rate is only $\approx 0.5\%$. 
More information on the SVX tag can be found in
Ref.~\cite{CDF_top_prd,CDF_top_disc}.

D\O\ uses kinematic
and topological cuts as well as $b$ tagging via soft muon tagging
to reduce
the background in the lepton plus jets channel. The
first approach exploits the fact that jets from $t\bar t$ decays tend to be
more energetic and more central than from typical QCD background
events. In addition, $t\bar t$ events are more spherical 
while QCD jet production results in more planar event shapes. 
Top enriched data samples can therefore be selected with a set of
topological and kinematic cuts like 
missing energy or aplanarity.
The second D\O\ approach uses $b$ tagging via muon tags  
through  $b \ra \mu X$ and $b \ra c \ra \mu X$ decays.
For more details on both
techniques see Ref.~\cite{D0_top_disc,D0_top_mass}.
The lepton plus jets event summary for
the CDF and D\O\ experiment is shown in Table~\ref{dilep_sum} (right).
 
\subsubsection{The Top All Hadronic Channel}
 
The signature of the all hadronic channel (Fig.~\ref{top_dec_sig1}c)
is nominally six jets where two of them are from $b$ quarks, no
leptons, and low \met. 
In order to overcome the huge background from QCD multijet
production, $b$ tagging as well as kinematic requirements are
used.
If the backgrounds can be controlled, the all hadronic channel would
be well suited to determine the top mass because all objects of the
top decay are measured in the detector. 

After kinematic cuts are applied in the CDF all hadronic
analysis~\cite{CDF_all_had}, at least five jets are required, where
the leading jets have to 
pass an aplanarity cut. In addition, at least one jet has to be SVX
tagged.  CDF observes 187 events on a predicted background of
$151\pm10$ events.
In the search for $t\bar t$ events in the all hadronic mode, D\O\ 
requires six jets, 
where a soft muon
tag has to be present in at least one of the jets. 
D\O\ combines several kinematic variables and uses a neural net
approach to separate signal from background obtaining 44 events with
an expected background of $25.3\pm7.3$ events.

\subsection{The Top Production Cross Section}

The measurement of the top production cross section is
given by 
$\sigtt=(N_{\rm obs}-N_{\rm bkg})/A\,{\cal L}$.
The number of predicted background events $N_{\rm bkg}$ is subtracted
from the number of observed top candidates $N_{\rm obs}$ and divided
by the acceptance $A$ of the sample selection and the integrated
luminosity $\cal L$ of the used data set. The measurement of \sigtt\
has been determined in several decay channels. The results of the
different \sigtt\ 
measurements from CDF~\cite{CDF_top_prod} are 
$8.2^{+4.4}_{-3.4}$~pb,
$6.7^{+2.0}_{-1.7}$~pb, and
$10.1^{+4.5}_{-3.6}$~pb
for the dilepton, lepton plus jets, and all hadronic mode,
respectively.
CDF measures a combined top production cross section of 
$\sigtt = 7.6^{+1.8}_{-1.5}$~pb. 
The results from the D\O\ analysis~\cite{D0_top_prod} are
$6.3\pm3.3$~pb from the dilepton and $e\nu$ channel, 
$4.1\pm2.0$~pb from lepton plus jets, and
$8.2\pm3.5$~pb from lepton plus jets with a $\mu$ tag.  
D\O\ quotes a combined top cross section of $\sigtt = (5.5\pm1.8)$~pb. 
Theoretical predictions~\cite{laenen,top_theory} range 
between 4.7--5.5~pb for a top quark mass of 175~\gevcc.

\subsection{The Top Quark Mass Measurement}

The top quark mass \mtop\ is a fundamental parameter of the Standard Model. A
precise determination of \mtop\ is one of the most
important measurement of CDF and D\O\ in Run~I. 
Because of the large background in the all hadronic mode and the low
number of top events in the dilepton channel, the most powerful 
dataset for measuring the top quark mass is the lepton
plus jets sample. 
The preferred method to determine \mtop\ 
is a constrained fit to the lepton plus 4-jet events
arising from the process $t\bar t \ra W b W \bar b \ra
\ell \nu b q \bar q^{\prime} \bar b$. The 
exact correspondence between the observed lepton, jets, and \met\ and
the $t\bar t$ decay products is not known.  
There are 12
combinations to assign the observed objects to the partons
from the $t\bar t$ decay.
These are reduced to six (two) combinations
if one (two) $b$~jets are tagged. Usually the mass fitter decides on a
preferred assignment based on a $\chi^2$ variable.

\begin{figure}[tbp]
\begin{picture}(160,70)(0,0)
\put(0,68){\large\bf (a)}
\put(85,68){\large\bf (b)}
\centerline{
\epsfysize=7.5cm
\epsffile[30 155 525 650]{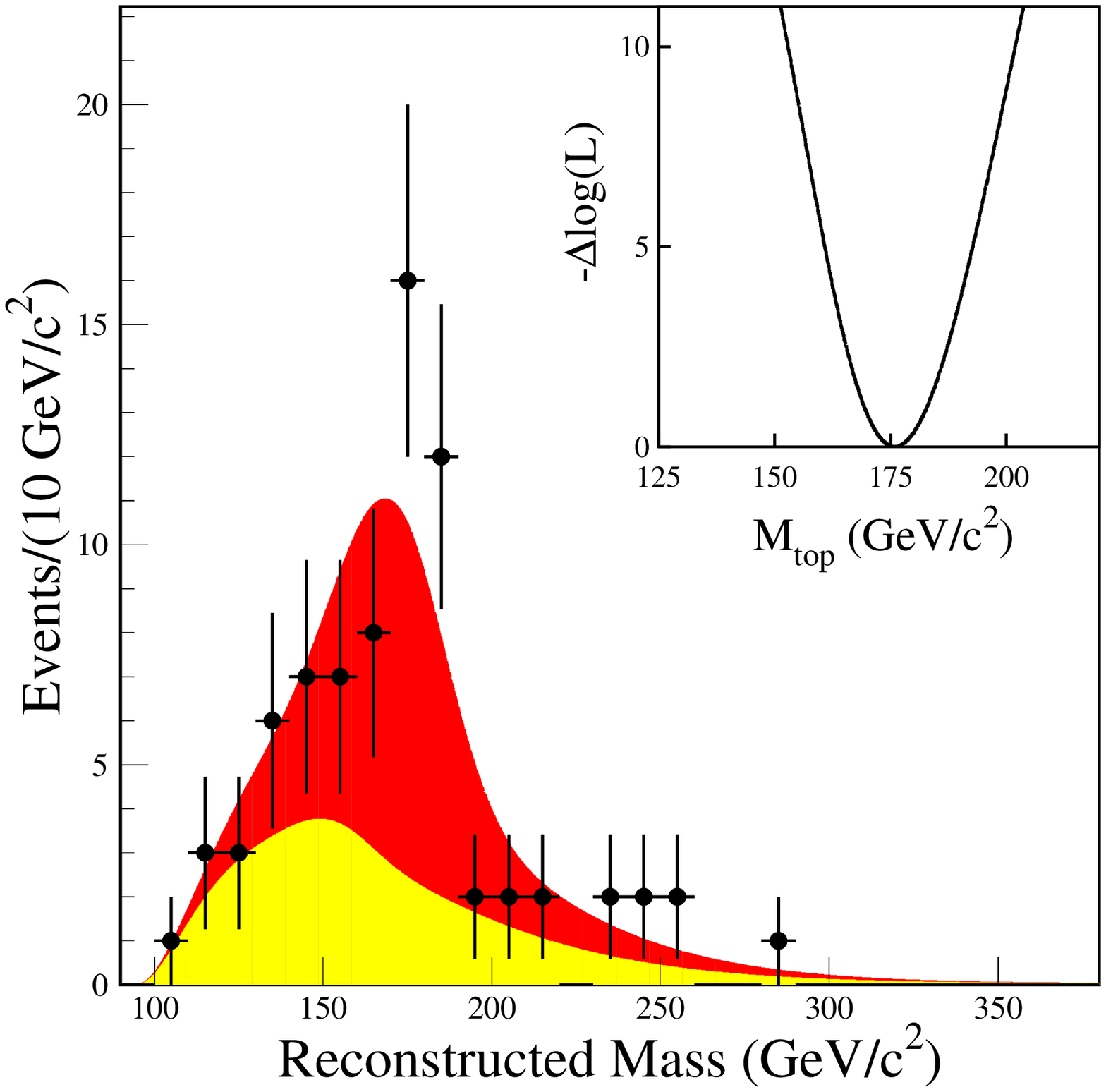}
\hspace*{1.0cm}
\epsfysize=7.5cm
\epsffile[120 175 490 620]{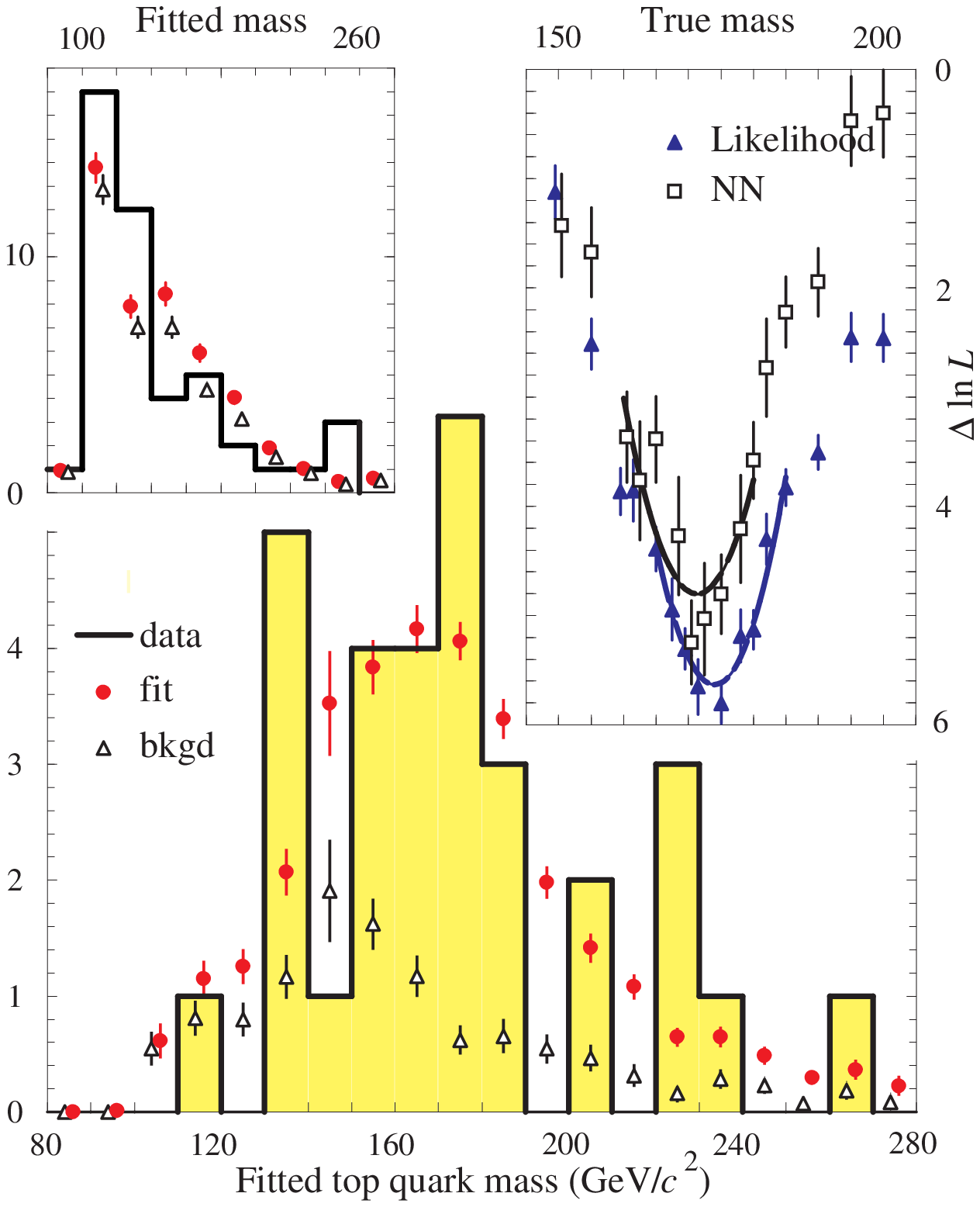}}
\end{picture}
\caption{Reconstructed mass distribution of (a) the CDF and (b) the
 D\O\ top quark mass 
 measurements in the lepton plus jets decay channel. See text for more
 information. }
\label{top_mass}
\end{figure}

At CDF the lepton plus 4-jet sample with at least one
SVX tag provided the original top mass measurement~\cite{CDF_top_disc}.
Optimization studies indicated a reduced error on \mtop, if the
tagged events are partitioned into non-overlapping tagging classes and a set
of untagged events is added. The following four data samples are used:
events with a single SVX tag (15 events), 
events with two SVX tags (5 events),
events with a SLT tag but no SVX tag (14 events), and events
with no tag but all four leading jets have $E_{\rm t}>15$~GeV (42
events). 
The reconstructed mass distribution of the sum of the four
subsamples is plotted in Fig.~\ref{top_mass}a). The data points are
compared with the result of the combined fit (dark shading) and the
fitted background component (light shading). The inset shows 
the shape of the log-likelihood as a function of \mtop. A value of 
$\mtop~=~(175.9\pm4.8\pm4.9)$~\gevcc\ has been extracted, where the
main systematic error is from the jet energy scale. Further
details on the CDF top quark mass measurement can be found in
Ref.~\cite{CDF_top_mass}. 

The D\O\ top mass measurement~\cite{D0_top_mass} is based on 77
lepton plus jets events, 
where five events are $\mu$ tagged and about 65\% are
background. Further separation of signal and background is based on
four kinematic variables, which are chosen to have small correlations
with \mtop. D\O\ also engages in a neural net approach, which is
sensitive to these kinematic variables as well as their correlations. 
The reconstructed top mass distribution of the final lepton plus jets sample
is shown in Fig.~\ref{top_mass}b). 
The shaded histogram are the data, while the solid circles represent the
predicted mixture of top and background, and the triangles are the
predicted background only. D\O\ extracts 
$\mtop~=~(173.3\pm5.6\pm6.2)$~\gevcc, where the
main systematic error also comes from the jet energy scale.

\begin{figure}[tbp]
\begin{picture}(160,68)(0,0)
\put(-2,65){\large\bf (a)}
\put(77,65){\large\bf (b)}
\centerline{
\epsfysize=7.2cm
\epsffile[60 150 580 680]{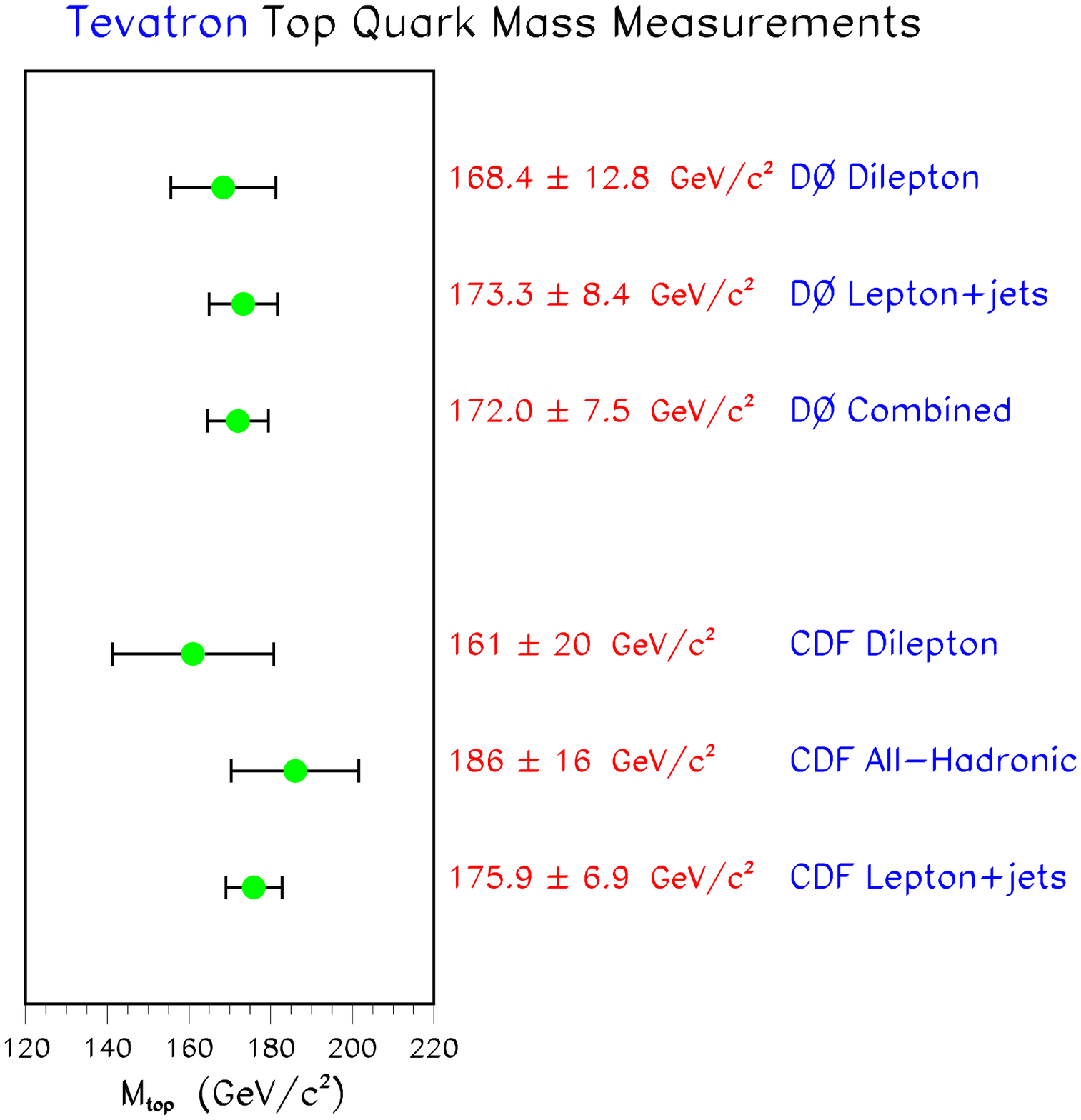}
\hspace*{0.5cm}
\epsfysize=7.2cm
\epsffile[70 180 525 625]{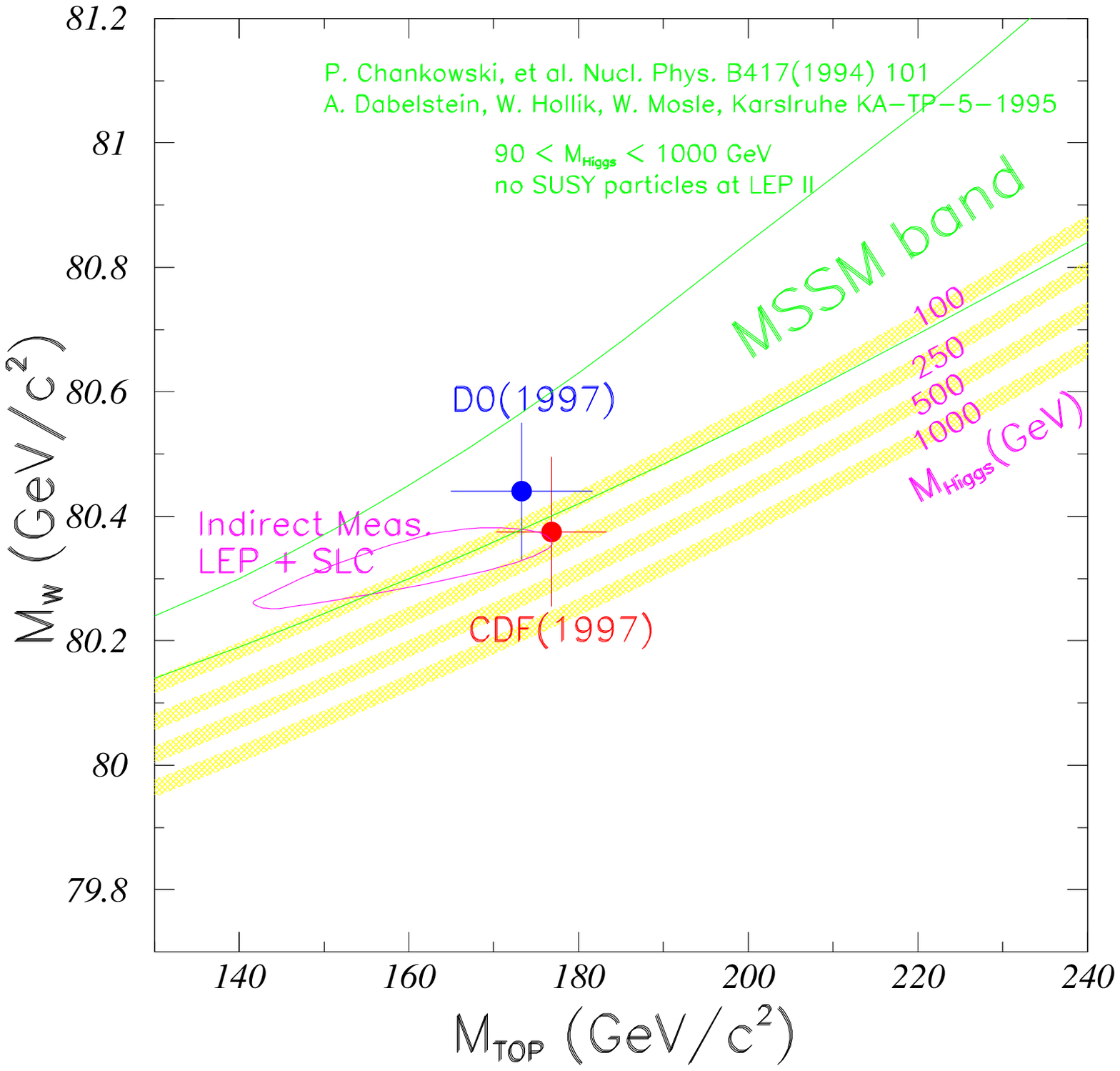}}
\end{picture}
\caption{(a) Summary of the CDF and D\O\ top quark mass
measurements in different decay channels. (b) Relation between the top
quark mass and the $W$ boson mass.}
\label{top_mass_sum}
\end{figure}

A summary of the top quark mass measurement by CDF and D\O\ is shown
in Figure~\ref{top_mass_sum}a). Both experiments also measured \mtop\ in
the dilepton channel, while CDF has a top mass determination from the
all hadronic channel~\cite{CDF_all_had} in addition. 

Since the top quark mass is
large, it controls the strength of quark loop corrections to
electroweak parameters like the $W$ boson mass $m_W$. If \mtop\ and
$m_W$ are precisely 
measured, the Standard Model Higgs boson mass can be constraint as
shown in Fig.~\ref{top_mass_sum}b). The Tevatron results as well
as the indirect measurements at the $Z$ pole seem to indicate a small
Higgs boson mass, but the uncertainties are still large and more data
is needed.

\section{Bottom Quark Physics at the Tevatron}
\label{b_sec}

The principal interest in studying $B$ hadrons 
in the context of the Standard Model 
arises from the fact that $B$ hadron decays provide valuable information
on the weak mixing matrix, the Cabibbo-Kobayashi-Maskawa (CKM)
matrix~\cite{ckm}. Traditionally, $B$ physics has been the domain of
$e^+ e^-$ machines, but 
already the UA\,1 collaboration has shown that $B$ physics
is feasible at a hadron collider \cite{bfeasi}. 
In this report, we concentrate 
on recent results on $B$~hadron lifetimes and time dependent
$B^0 \bar B^0$ oscillations where there exist only results from the
CDF experiment.

The main production mechanism of $b$ quarks at the Tevatron is through
gluon-gluon fusion. Compared to top quark production, 
the $b$ quark production cross
section is quite large with $\sigma_{b} \sim 50~\mu$b within the
central detector region of
rapidity less than one. This huge cross section 
resulted in about $5\cdot 10^9\ b\bar b$ pairs being produced in
Run~I.
But the total
inelastic cross section is still about three orders of magnitude
larger. This puts 
certain requirements on the trigger system in finding $B$ decay
products. 
Because of the rapidly
falling $b$ production cross section,
one likes to go as low as possible in the \Pt\ trigger thresholds to
increase the amount of 
recorded $B$ triggers with
the DAQ bandwidth being the limiting factor.
All $B$ physics triggers at CDF are based on leptons.
Dilepton triggers with principal \Pt\ thresholds of about 2 \gevc\ 
per lepton as well as
single lepton triggers with \Pt\ thresholds around 7.5 \gevc\ both exist.
An additional
basis of CDF's $B$ physics program are the good tracking and
vertexing capabilities of the CTC and SVX. 

\subsection{{\boldmath $B$} Hadron Lifetime Measurements}

\begin{figure}[tbp]
\begin{picture}(160,70)(0,0)
\put(10,65){\large\bf (a)}
\put(93,65){\large\bf (b)}
\centerline{
\epsfysize=7.5cm
\epsffile{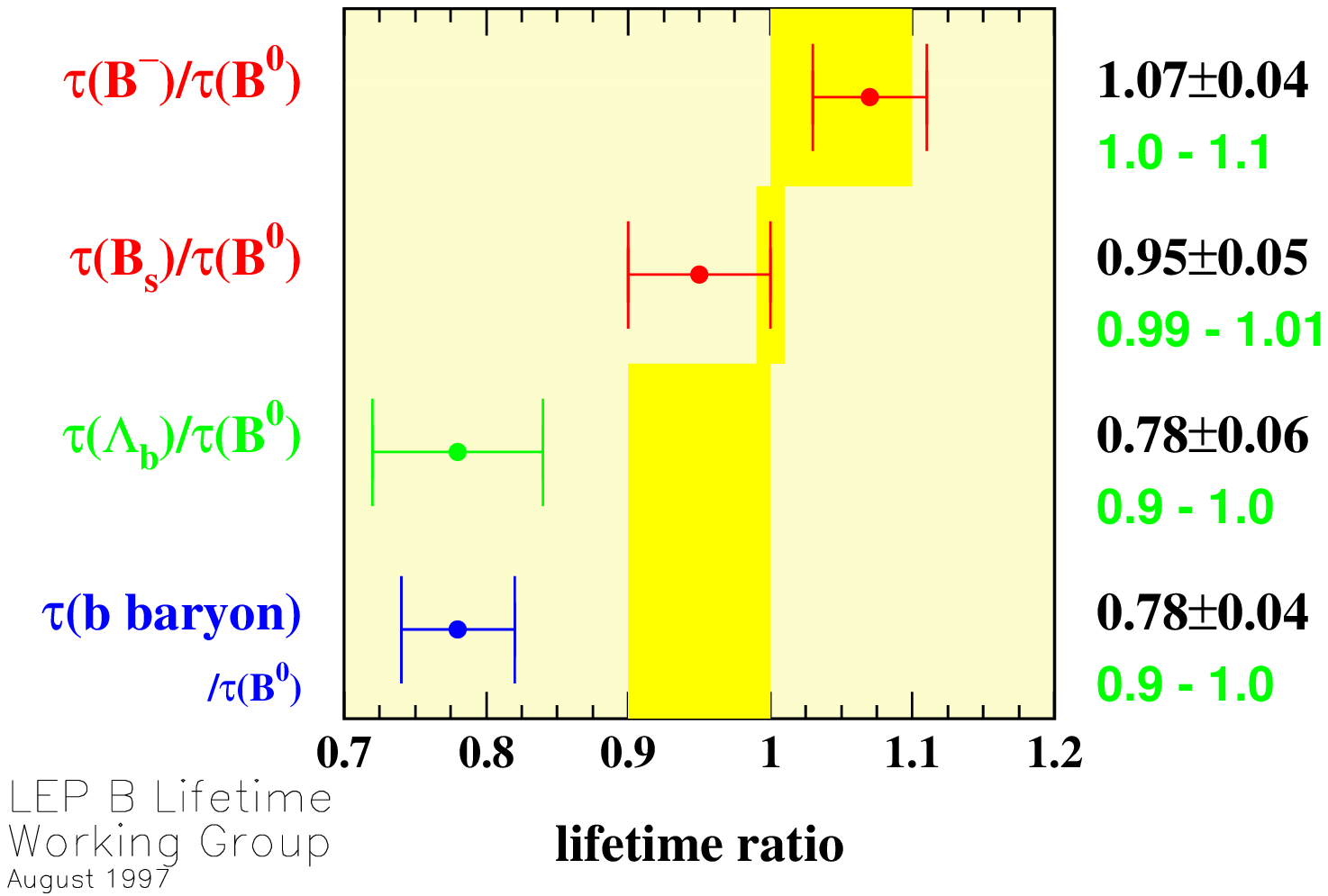}
\hspace*{-0.5cm}
\epsfysize=7.5cm
\epsffile[5 45 550 630]{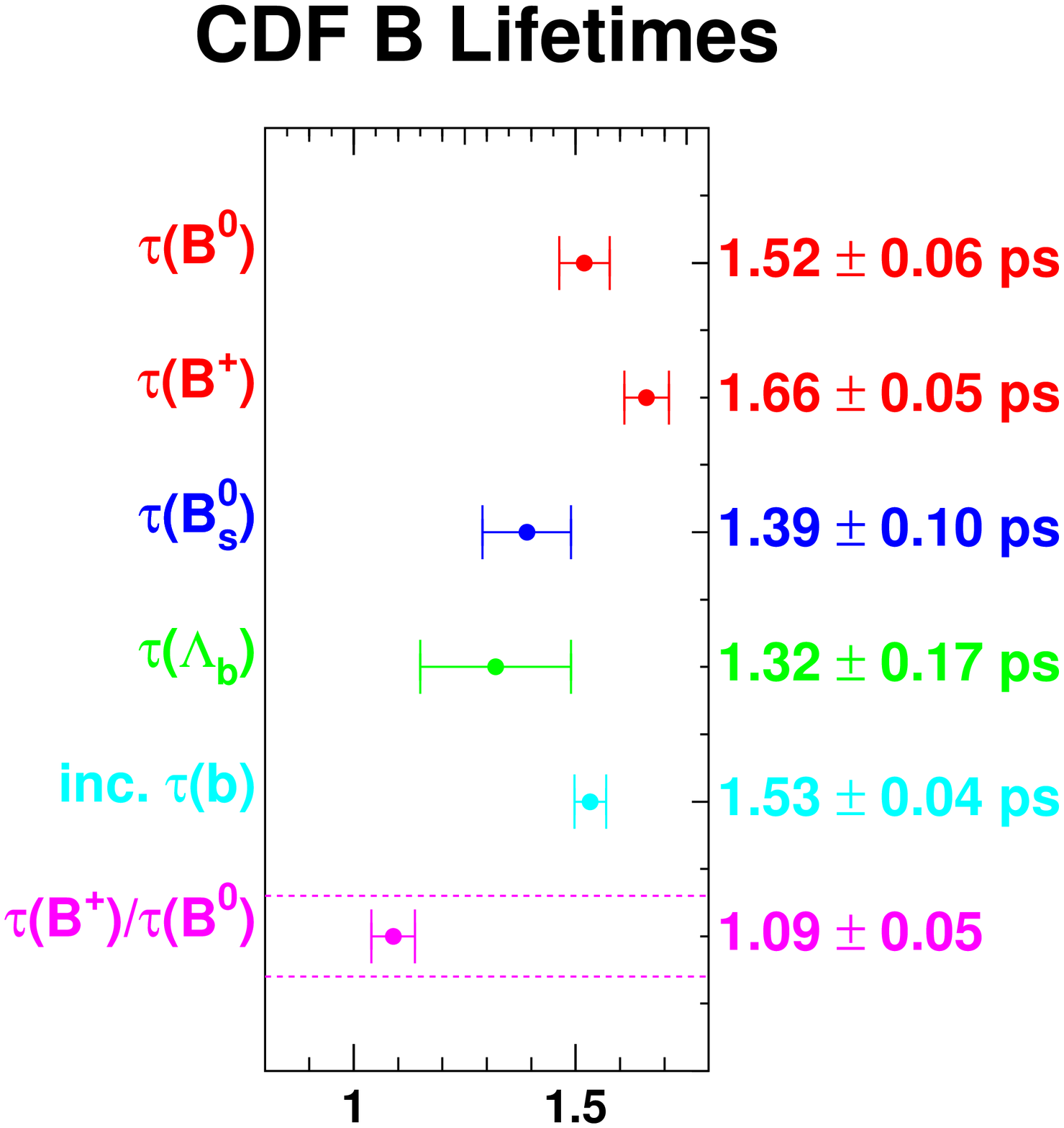}}
\end{picture}
\caption{Summary of (a) $B$ lifetime ratios and (b) CDF 
$B$ hadron lifetime measurements.}
\label{blife_sum}
\end{figure}

The lifetime differences between bottom flavoured hadrons
can probe the $B$~decay  
mechanisms which are beyond the simple quark spectator model. 
In the case of charm mesons, such differences have been observed to 
be quite large ($\tau(D^+)/\tau(D^0)\sim 2.5$) \cite{PDG}.
Among bottom hadrons, the lifetime differences are expected to be 
smaller due to the heavier bottom quark mass \cite{bigi,neubert}. 
Some phenomenological models \cite{bigi} predict a lifetime difference between 
the $B^+$ and $B^0$ meson of about 5\% 
and between the $B^0$ and \Bs\ meson of about 1\%. A compilation
of different $B$ hadron lifetime ratios as determined by the LEP $B$
Lifetime Working Group are shown in Fig.~\ref{blife_sum}a).
The range of theoretical predictions are indicated as shaded bands in
this Figure.

CDF has measured the lifetimes of all $B$ hadrons by fully or
partially reconstructing $B^0$, $B^+$, and \Bs\ mesons as well as
$\Lambda_b$ baryons. A compilation of the 
precise CDF $B$~hadron lifetime measurements can be found in
Fig.~\ref{blife_sum}b). In the following, 
we will describe the \Bs~lifetime measurement using \Dsl\
correlations, which has recently been updated. 
 
\subsubsection{{\boldmath \Bs} Meson Lifetime Measurement}   
\label{bs_life_sec}

The lifetime of the \Bs\ meson is measured at CDF using the semileptonic decay 
$\Bs \ra \Dsl \nu X$, where the \Ds\ is reconstructed through its
decay modes into \phipi, \kstark, \ksk, and \phil.  
For the first 3 decay modes the analysis
starts with a single lepton trigger data set, while the
semileptonic \Ds\ decay mode is based on a dimuon data sample obtained
with a trigger requirement of $m(\mu\mu) < 2.8$~\gevcc. 
\Ds\ candidates are searched in a cone around the lepton and then   
intersected with the lepton to find the \Bs\
decay vertex.
Since the \Bs\ meson is not fully
reconstructed, its \ctau\ cannot be directly determined. A correction
has to be applied to scale from the \Dsl\ 
momentum to $\Pt(\Bs)$. This $\beta\gamma$ correction is
obtained from a Monte Carlo simulation.   

\begin{figure}[tbp]
\begin{picture}(160,70)(0,0)
\put(18,63){\large\bf (a)}
\put(95,63){\large\bf (b)}
\centerline{
\epsfysize=7.5cm
\epsffile[5 5 550 540]{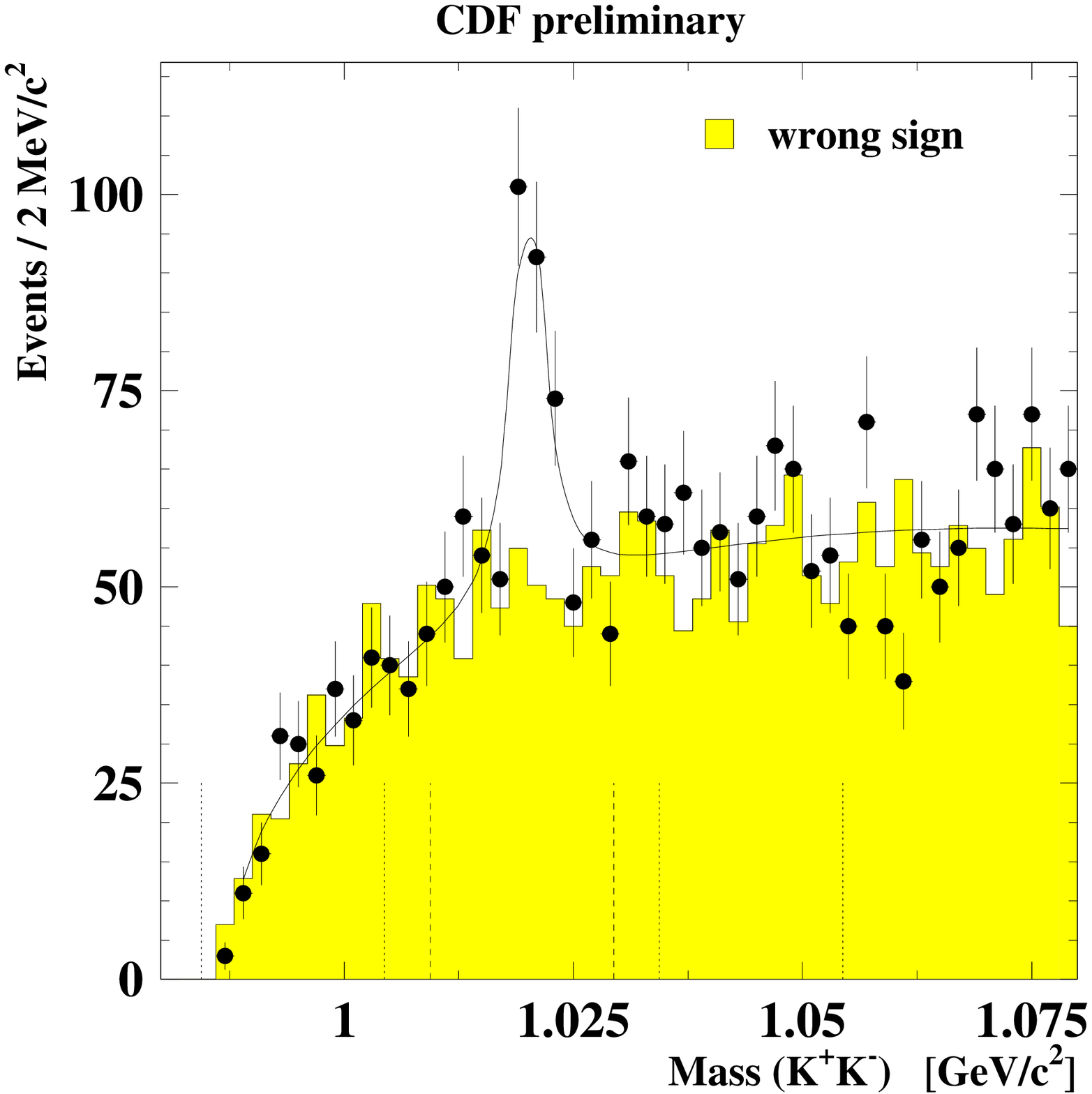}
\epsfysize=7.5cm
\epsffile[5 5 550 540]{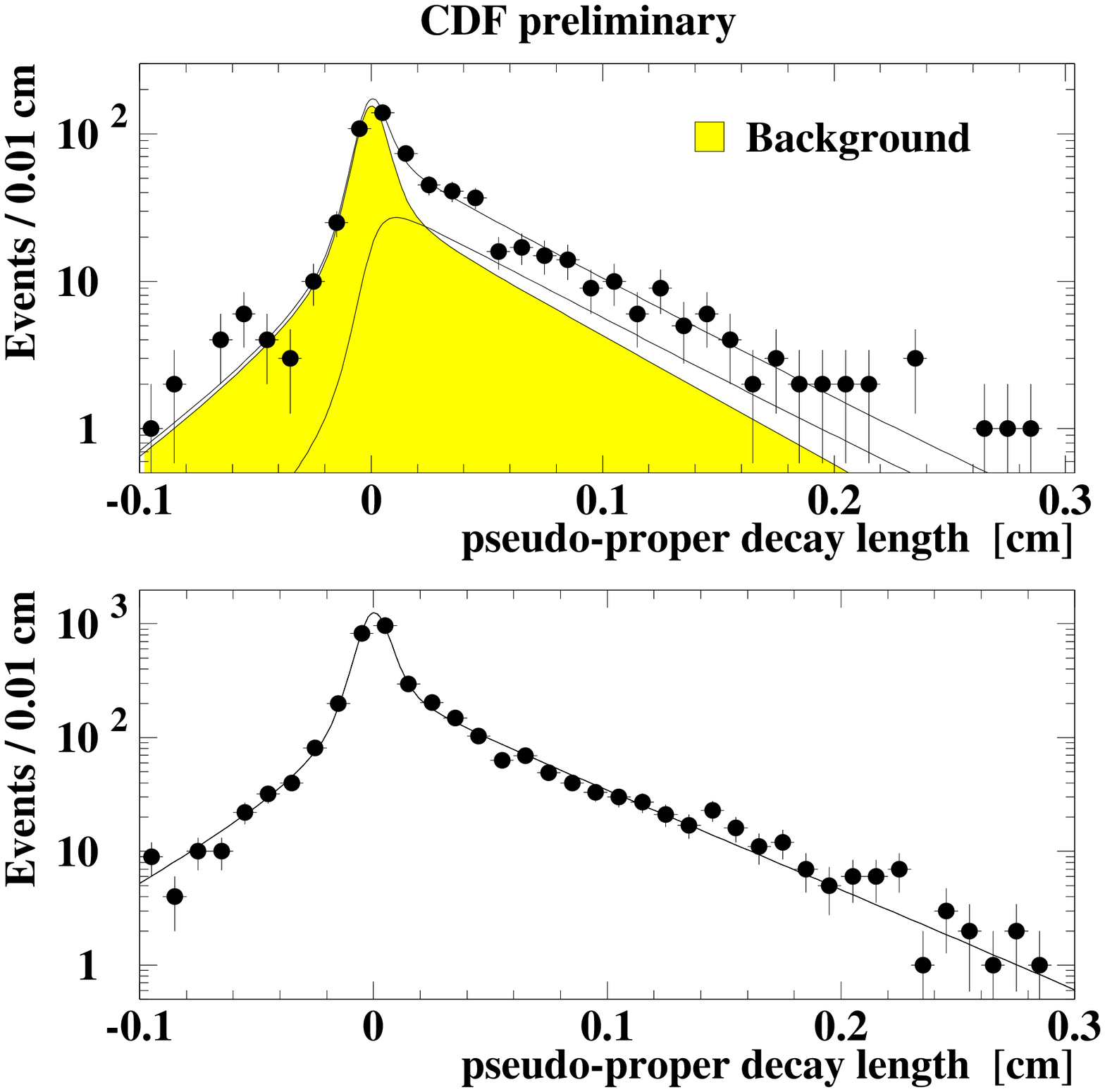}
}
\end{picture}
\caption{(a) Invariant $K^+K^-$ mass distribution of 
$\Bs \ra \Dsmu \nu X$ decays with $\Ds \ra \phil$, $\phi \ra K^+K^-$.
(b) Decay length distributions for signal (top) and 
background (bottom) in the $\Ds \ra \phil$ sample.} 
\label{bs_life}
\end{figure}

About 600 \Bs\ candidates have been reconstructed in the four \Ds\
decay channels, where the $\Ds \ra \phipi$ mode contributes the
largest statistics with $220\pm21$ events. As an example, the $K^+K^-$
invariant mass distribution from the $\Ds \ra \phil$ decay mode with
$\phi \ra K^+K^-$ is
shown in Fig.~\ref{bs_life}a) displaying $205\pm38$ signal events. 
Figure~\ref{bs_life}b) shows the decay length
distribution of the \phil\ mode, where the background underneath the
$\phi$ signal was obtained from the $\phi$ sidebands 
as displayed at the bottom of Fig.~\ref{bs_life}b).
From all four \Ds\ decay modes a \Bs\ lifetime 
$\tau(\Bs)=(1.39\pm0.09\pm0.05)$~ps has been
measured. The main systematic errors arise from the background shape
and from non-\Bs\ backgrounds.

In the Standard Model of the CKM mixing matrix, the ratio
$\dgam/\dm$ contains no CKM matrix elements and depends 
only on QCD corrections. If the error on this QCD calculation is
understood and not too    
large, a measurement of \dgam\ in the \Bs\ meson system implies a determination of \dms\ and thus a
way to infer $\Bs\bar\Bs$ mixing.
The \Bs\ meson proper decay length distribution has been examined
for a lifetime difference \dgog\ between the two $CP$ eigenstates of the
\Bs\ meson, \Bsh\ and \Bsl. 
Instead of fitting for a functional form of $\Gamma e^{-\Gamma t}$ the
likelihood fit has been expanded to fit for a functional form
$1/2\cdot(\Gamma_L e^{-\Gamma_L t} + \Gamma_H e^{-\Gamma_H t} )$
with $\Gamma_{L,H} = \Gamma \pm \Delta\Gamma/2$.
Fixing the average \Bs\ lifetime to the PDG value $\tau(\Bs) =
(1.57\pm0.08)$~ps~\cite{PDG}, the fit returns $\dgog = 0.48
^{+0.26}_{-0.48}$ indicating that the current statistics is not
sensitive to a \Bs\ lifetime difference. Based on this fit result a
limit on 
$\dgog < 0.81$ (95\% CL)
can be set. Using a value of
$\dgam/\dm = (5.6\pm2.6)\cdot 10^{-3}$
from~\cite{Buchalla}, an upper limit on the \Bs\ mixing frequency
\dms\ can be obtained 
$$\dms < 92 \times (5.6\cdot10^{-3})/(\dgam/\dm) \times 
 (1.57\,{\rm ps}/\tau_{\Bs})\ \ \ \ \ (95\%\ {\rm CL}).$$

\subsection{{\boldmath $B^0\bar B^0$} Oscillations}

In the Standard Model, $B^0\bar B^0$ mixing occurs through a second
order weak 
process, where the dominant contribution is through a top quark loop.
The oscillation is expressed in terms of its frequency $\dm_d$, where
$\Delta m_d$ is the 
difference in mass between the two $B^0$ meson eigenstates.
For a beam initially pure in $B^0$ mesons at $t=0$, the number of
$B^0$ that oscillate at proper time $t$ is given by
$
N(t)_{B^0 \rightarrow \bar{B}^0} 
        = 1/2\,\Gamma \exp(-\Gamma t)\cdot(1-\cos\Delta m_d\, t).
$
Measurements of the frequencies of $B^0$ and \Bs\ oscillations can
constrain the magnitudes of the CKM matrix elements
$V_{td}$ and $V_{ts}$.

\begin{figure}[tbp]
\begin{picture}(160,65)(0,0)
\put(-1,64){\large\bf (a)}
\put(78,64){\large\bf (b)}
\centerline{
\epsfysize=7.1cm
\epsffile[10 50 625 635]{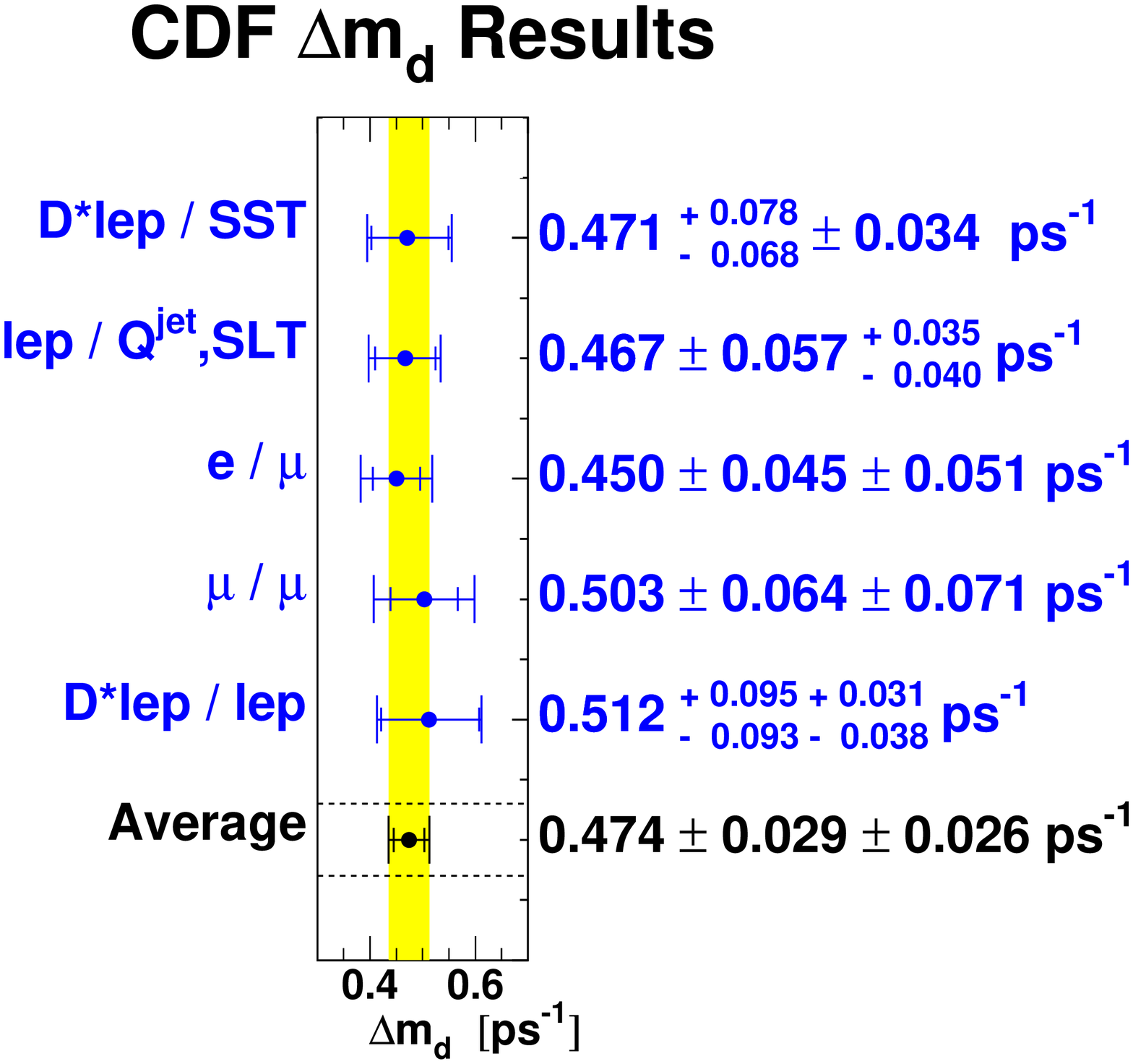}
\hspace*{0.7cm}
\epsfysize=7.1cm
\epsffile[5 5 530 510]{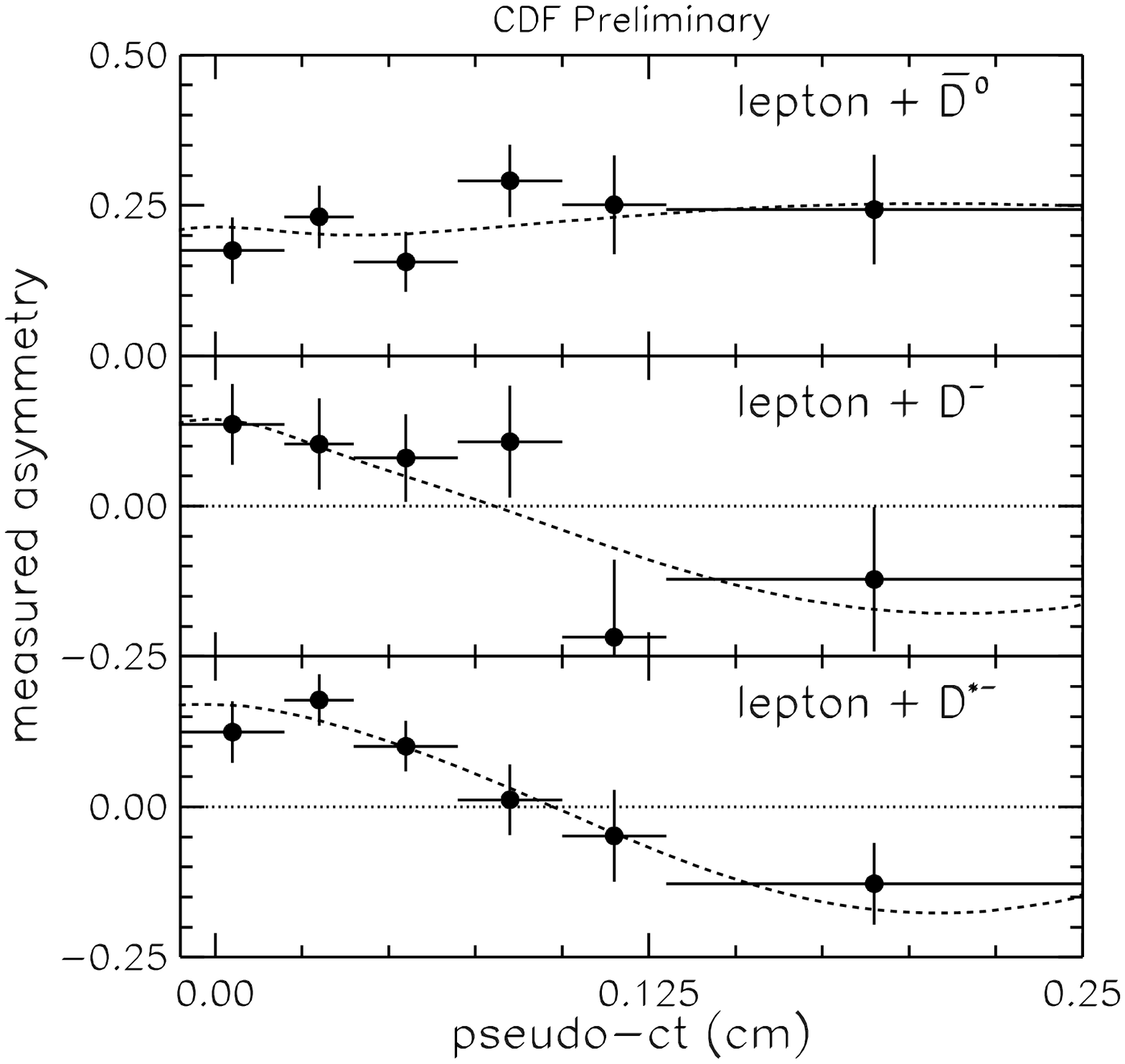}
}
\end{picture}
\caption{(a) Summary of time dependent $\dm_d$ results from CDF. (b) Measured
asymmetry from the $D^{(*)}\ell$ mixing
analysis.}
\label{bmix_sst}
\end{figure}

In general, a time dependent mixing measurement requires the knowledge
of the flavour of the $B$ meson at production and at decay, as well as
the proper decay time of the $B$~meson.
Experimentally, the flavour of the $B$~meson at decay time is determined 
from the observed decay products like the charge of
the lepton from a semileptonic $B$ decay.  The flavour
at production time can be determined in various ways, employing
either the second $b$-flavoured hadron in the event, or the charge
correlation with particles produced in association with the
$B^0$~meson (`same side tagging' SST). 
CDF has preliminary results on several time-dependent $B^0$ mixing
analyses utilizing several ways to tag the $B$ flavour at production.
These measurements include opposite side lepton tagging, jet
charge tagging, or SST as can be seen in Fig.~\ref{bmix_sst}a).
The combined average from the CDF $\dm_d$ measurements is 
$\dm_d=(0.474\pm0.029\pm0.026)$~ps$^{-1}$, which is competitive with
results from other experiments~\cite{PDG}.
In the following, we report 
about one of CDF's $\dm_d$ measurements exploiting the same
side tag, which is of particular interest at a hadron collider.

\subsubsection{{\boldmath $B^0\bar B^0$} Mixing in {\boldmath 
$D^{(*)}\ell$} Events} 

For this analysis, further described in Ref.~\cite{bmix_sst}, $B$~mesons
are reconstructed through  
their semileptonic decays $B \ra D^{(*)}\ell\nu X$. The analysis starts with
single lepton trigger data and reconstructs $D^{(*)}$ meson
candidates in a cone around the trigger electron or muon in the
following channels: 
$\bar B^0 \ra D^{*+}\ell^-\nu$,
$D^{*+} \ra D^0 \pi^+$ with
$D^0 \ra K^- \pi^+$,
$K^- \pi^+\pi^+\pi^-$, and
$K^- \pi^+ \pi^0$ ($\pi^0$ not  reconstructed) as well as
$\bar B^0 \ra D^+\ell^-\nu$, 
$D^+ \ra K^- \pi^+ \pi^+$, and
$B^- \ra D^0\ell^-\nu$, with
$D^0 \ra K^- \pi^+$ (veto $D^{*+}$).
About 7000 partially reconstructed $B^0$~mesons and about 2000
$B^+$ candidates are selected this way. 
The $D^{(*)}$ candidates are intersected with the
lepton to find the $B$ decay vertex
in a similar way as described 
in Sec.~\ref{bs_life_sec}.

To tag the $B$ flavour at production, a same side
tagging algorithm, which exploits the correlation between the
$B$ flavour and the charge of tracks from either the fragmentation
process or $B^{**}$ resonances~\cite{Rosner}, is used.
In this analysis no attempt is made to
differentiate the sources of correlated pions.
To study the correlation between the flavour of the $B$~meson and
the charged particles produced in association with it, 
all tracks within a $\eta$-$\phi$ cone of radius $0.7$ centered
around the direction of the $D^{(*)}\ell$ candidate are used.
The tracks considered as tags should be consistent with the hypothesis
that they originate from the fragmentation chain or the decay of
$B^{**}$ mesons; this means they are required to come from the
primary event vertex.

String fragmentation models indicate that the velocity of the fragmentation
particles is close to the velocity of the $B$ meson.
Similarly, pions from $B^{**}$ decays should also have a velocity 
close to the one of the $B$ meson.  In particular, the relative transverse
momentum ($\ptrel$) of the particle with respect to
the combined momentum of the $D^{(*)}\ell$ combination plus tag
particle momentum, 
should be small. 
Of the candidate tracks, we select as the tag the track that has the minimum
component of momentum $\ptrel$ to the momentum sum of that track,
and the $D^{(*)}\ell$~combination.
The efficiency for finding such a tag is $\sim70\%$.

Since we know the flavour of the $B$ meson at decay from the
$D^{(*)}\ell$ signature, we compare the number of right-sign ($N_{RS}$)
SST tags to the number of wrong-sign ($N_{WS}$) tags as a function
of $c\tau$. For the $B^0$ meson we expect the asymmetry $A(t)$ to be
$
A(t) = (N_{RS}(t) - N_{WS}(t))/(N_{RS}(t) + N_{WS}(t)) =
	D\cdot\cos(\Delta m_d\,t),
$
where $D$ is the dilution of the same side tagging algorithm. $D$ is also
related to the probability $w$ of mistagging the flavour 
by $D = 1 - 2\, w$.
In this analysis both $\Delta m_d$ and $D$ are determined simultaneously.

To obtain the asymmetry for $B^0$ and $B^+$ mesons, a correction is applied
for the fact that each $D^{(*)}\ell$ signal has contributions from both
neutral and charged $B$ mesons via $D^{**}$ decays.
The fit result is shown in Fig.~\ref{bmix_sst}b). The fit yields 
$
\Delta m_d = (0.471^{+0.078}_{-0.068}\pm0.034)\ {\rm ps}^{-1}, 
$
as well as the neutral and charged dilutions 
$D_0 = 0.18\pm0.03\pm0.02$ and $D_+ = 0.27\pm0.03\pm0.02$, respectively. 

\section{Conclusions}
\label{conclude_sec}

In this article, we have reviewed recent heavy quark physics results from
the Tevatron $p\bar p$~collider at Fermilab.
We summarized the status of top quark physics at
CDF and D\O\ including measurements of
the top production cross section 
and the top quark mass in several top decay modes.
We also discussed 
recent $B$ physics results from the CDF experiment including
$B$ hadron lifetime measurements, which are very competitive with the
LEP and SLC results, as well as time dependent $B^0\bar B^0$ mixing.
These results give rise to good
prospects of top and $B$ physics at the Tevatron 
in Run~II starting in 1999.

\subsubsection*{Acknowledgments}

It is a pleasure to thank all friends and colleagues from the CDF
and D\O\ collaboration for their excellent work and their help in
preparing this talk. 
Special thanks go to M.~Beyer, T.~Mannel and H.~Schr\"oder for
organizing such a delightful workshop.
A constant source of inspiration and support is my wife Ann,
who I like to thank for her love and her 
continuous understanding about the life of a physicist.

\end{document}